%% file: main.tex
\documentclass[aps,prl,reprint,superscriptaddress,numbers,sort&compress]{revtex4-2}
\usepackage{blindtext}
\RequirePackage[T1]{fontenc}
\RequirePackage{fix-cm}

\usepackage[american]{babel}

\RequirePackage{graphicx}
\RequirePackage{mathptmx}
\usepackage{comment}

\DeclareMathAlphabet{\mathcal}{OMS}{cmsy}{m}{n}

\usepackage{footmisc}
\RequirePackage{natbib}
\RequirePackage{siunitx}
\RequirePackage{amsmath}
\usepackage{caption}
\usepackage{upgreek}
\usepackage{subfig}
\usepackage{caption}
\usepackage{float}
\usepackage{ragged2e}
\usepackage{bbold}

\usepackage{bm}

\graphicspath{{./Images/}}

\usepackage{url}
\RequirePackage[colorlinks,citecolor=blue,urlcolor=blue,linkcolor=blue]{hyperref}

\begin{document}

\raggedbottom

\title{New high-sensitivity search for neutron to mirror-neutron oscillations at the PSI UCN source}

\author{N.~J.~Ayres}
\affiliation{Institute for Particle Physics and Astrophysics, ETH Zürich, 8093 Zürich, Switzerland}
% agreed to authorship 

\author{Z.~Berezhiani}
\affiliation{INFN, Laboratori Nazionali del Gran Sasso, Assergi, 67100 L’Aquila, Italy}
% agreed to authorship

\author{G.~Bison}
\affiliation{Laboratory for Particle Physics, PSI Center for Neutron and Muon Sciences, Paul Scherrer Institute (PSI), 5232 Villigen, Switzerland}

\author{K.~Bodek}
\affiliation{Marian Smoluchowski Institute of Physics, Jagiellonian University, 30-348 Cracow, Poland}
% agreed to authorship

\author{V.~Bondar}
\affiliation{Institute for Particle Physics and Astrophysics, ETH Zürich, 8093 Zürich, Switzerland} 
% agreed to authorship 

\author{P.-J. Chiu}
\altaffiliation[Present address: ]{Department of Physics, National Taiwan University, 106319 Taipei, Taiwan}
\affiliation{Institute for Particle Physics and Astrophysics, ETH Zürich, 8093 Zürich, Switzerland}
\affiliation{Laboratory for Particle Physics, PSI Center for Neutron and Muon Sciences, Paul Scherrer Institute (PSI), 5232 Villigen, Switzerland}
% agreed to authorship

\author{M.~Daum}
\affiliation{Laboratory for Particle Physics, PSI Center for Neutron and Muon Sciences, Paul Scherrer Institute (PSI), 5232 Villigen, Switzerland}
% agreed to authorship 

\author{C.~B.~Doorenbos}
\affiliation{Institute for Particle Physics and Astrophysics, ETH Zürich, 8093 Zürich, Switzerland}
\affiliation{Laboratory for Particle Physics, PSI Center for Neutron and Muon Sciences, Paul Scherrer Institute (PSI), 5232 Villigen, Switzerland}
% agreed to authorship 

\author{S.~Emmenegger}
\affiliation{Institute for Particle Physics and Astrophysics, ETH Zürich, 8093 Zürich, Switzerland}
% agreed to authorship

\author{K.~Kirch}
\affiliation{Institute for Particle Physics and Astrophysics, ETH Zürich, 8093 Zürich, Switzerland}
\affiliation{Laboratory for Particle Physics, PSI Center for Neutron and Muon Sciences, Paul Scherrer Institute (PSI), 5232 Villigen, Switzerland}
% agreed to authorship 

\author{V.~Kletzl}
\altaffiliation[Present address: ]{Marietta-Blau-Institute for Particle Physics, Austrian Academy of Sciences, 1010 Vienna, Austria}
\affiliation{Institute for Particle Physics and Astrophysics, ETH Zürich, 8093 Zürich, Switzerland}
\affiliation{Laboratory for Particle Physics, PSI Center for Neutron and Muon Sciences, Paul Scherrer Institute (PSI), 5232 Villigen, Switzerland}
% Victoria.Kletzl-Teuffenbach@oeaw.ac.at

\author{J.~Krempel}
\affiliation{Institute for Particle Physics and Astrophysics, ETH Zürich, 8093 Zürich, Switzerland}
% agreed to authorship 

\author{B.~Lauss}
\email[Corresponding author: ]{bernhard.lauss@psi.ch}
\affiliation{Laboratory for Particle Physics, PSI Center for Neutron and Muon Sciences, Paul Scherrer Institute (PSI), 5232 Villigen, Switzerland}
% agreed to authorship 

\author{D.~Pais}
\affiliation{Institute for Particle Physics and Astrophysics, ETH Zürich, 8093 Zürich, Switzerland} 
\affiliation{Laboratory for Particle Physics, PSI Center for Neutron and Muon Sciences, Paul Scherrer Institute (PSI), 5232 Villigen, Switzerland}
% agreed to authorship

\author{I.~Rienäcker}
\affiliation{Laboratory for Particle Physics, PSI Center for Neutron and Muon Sciences, Paul Scherrer Institute (PSI), 5232 Villigen, Switzerland}
% ingo.rienaecker@gmail.com

\author{D.~Ries}
\affiliation{Laboratory for Particle Physics, PSI Center for Neutron and Muon Sciences, Paul Scherrer Institute (PSI), 5232 Villigen, Switzerland}
% agreed to authorship

\author{D.~Rozpędzik}
\affiliation{Marian Smoluchowski Institute of Physics, Jagiellonian University, 30-348 Cracow, Poland}
% agreed to authorship 

\author{P.~Schmidt-Wellenburg}
\affiliation{Laboratory for Particle Physics, PSI Center for Neutron and Muon Sciences, Paul Scherrer Institute (PSI), 5232 Villigen, Switzerland}
% agreed to authorship 

\author{K.~S.~Tanaka}
\altaffiliation[Present address: ]{Waseda University, 3-4-1 Ookubo, Shinjuku-ku, Tokyo 169-8555, Japan}
\affiliation{Laboratory for Particle Physics, PSI Center for Neutron and Muon Sciences, Paul Scherrer Institute (PSI), 5232 Villigen, Switzerland}
% agreed to authorship 

\author{J.~Zejma}
\affiliation{Marian Smoluchowski Institute of Physics, Jagiellonian University, 30-348 Cracow, Poland}
% agreed to authorship 

\author{N.~Ziehl}
\email[Corresponding author: ]{ziehln@phys.ethz.ch}
\affiliation{Institute for Particle Physics and Astrophysics, ETH Zürich, 8093 Zürich, Switzerland}
% agreed to authorship 

\author{G.~Zsigmond}
\email[Corresponding author: ]{geza.zsigmond@psi.ch}
\affiliation{Laboratory for Particle Physics, PSI Center for Neutron and Muon Sciences, Paul Scherrer Institute (PSI), 5232 Villigen, Switzerland}
% agreed to authorship 

\begin{abstract}
In a search for potential signals of neutron $(n)$ to mirror-neutron $(n')$ oscillations, a collaboration centered at the Paul Scherrer Institute (PSI) investigated the remaining parameter space claimed by anomalies with a dedicated high-sensitivity apparatus. 
An elaborate magnetic-field-mapping analysis and Monte Carlo simulation of the cumulative $n-n'$ oscillation probabilities along the neutron trajectories inside the storage vessel complemented the neutron data analysis. 
Magnetic fields were scanned in the range $\SI{5}{\micro\tesla} < B < \SI{109}{\micro\tesla}$. No evidence of anomalous neutron losses was found. Consequently, new limits for the $n-n'$ oscillation time constant were set. The parameter space, previously claimed for potential signals, has been  excluded to 99.98~\%.

\end{abstract}

\maketitle

%\section{Introduction}
\label{sec:intro}
\input{introduction.tex}

%\section{Experiment}
\label{sec:Experiment}
\input{experiment.tex}

%\section{Analysis}
\label{sec:limits}
\input{analysis.tex}

%\section{Conclusions}
%\label{sec:Conclusion}
%\input{conclusion.tex}

%\section{Acknowledgements}

\medskip 

\textit{Acknowledgments} $-$
The experiment would not have been possible without the 
excellent technical support of Michael Meier and Luke Noorda.
The authors acknowledge the valuable assistance of many support groups at PSI, especially the BSQ group operating the UCN source, the accelerator operating crews, the ‘Hallendienst’ and the mechanical workshops. 
We are also grateful for support from the ETH IPA and D-PHYS mechanical workshops and the vocational training division. 

ETH and PSI appreciate the financial support from the Swiss National Science Foundation through projects 162574 (ETH), 169596 (PSI), 172626 (PSI), 172639 (ETH), 178951 (PSI), 188700 (PSI), 196416 (ETH), 200441 (ETH), 10003932 (ETH). 
We further acknowledge funding from the ETH Career Seed Grant SEED-13 20-2 and the SNF spark programme grant CRSK-2\_196416. 
The collaborators from the Jagiellonian University Cracow wish to acknowledge support from the National Science Center, Poland, under grant No. 2016/23/D/ST2/00715, No. 2018/30/M/ST2/00319 and No. 2020/37/B/ST2/02349, and also by the Minister of Education and Science under the agreement No. 2022/WK/07.
The work of Z.B. was supported in part by the research grant No.
2022E2J4RK ``PANTHEON: Perspectives in Astroparticle and
Neutrino THEory with Old and New messengers" under the program
PRIN 2022 funded by the Italian Ministero dell'Universit\`a e della
Ricerca (MUR) and by the European Union – Next Generation EU.

%\clearpage

\bibliographystyle{apsrev4-2}

\def\urlprefix{}
%\newcommand{\doi}[1]{doi: \href{https://doi.org/#1}{#1}}
%\newcommand{\doi}[1]{doi: \href{https://doi.org/#1}{\nolinkurl{#1}}}
%\printbibliography
%\bibliography{Ref-short}
\bibliography{Zotero}
%\bibliographystyle{alpha}
%\bibliography{sample}
\newpage
\appendix

\input{appendix}
\end{document}

%% file: introduction.tex
%\subsection{Motivation}
A parallel mirror sector of particles as an exact duplicate of the ordinary particle sector was proposed a long time ago motivated by the issues of parity \cite{Lee:1956qn,Kobzarev:1966qya,Blinnikov:1983gh,Volkas2003}. 
In this picture all ordinary particles: the electron $e$, proton $p$, neutron $n$ etc., have mass degenerate mirror partners: $e'$, $p'$, $n'$ etc. which are sterile to our Standard Model interactions. Besides gravity, mirror particles can interact with ordinary matter only via specific feeble interactions \cite{Berezhiani:2003xm,Foot:2003eq}.  Mirror matter could account for dark matter, provided that during the evolution of the Universe it had a lower temperature than the ordinary sector~\cite{Berezhiani:2000gw,Ignatiev:2003js,Berezhiani:2003wj,Berezhiani:2005vv}.

Any neutral particle, elementary as e.g. neutrinos or composite as the neutron, can undergo a mixing  with its mass-degenerate mirror partner. For example, ordinary-mirror (active-sterile) neutrino oscillations were discussed in \cite{Foot:1995pa,Berezhiani:1995yi,Akhmedov:1992hh}. Interestingly, interactions that induce such mixing violate $B-L$ in each sector. Thus, in the light of Sakharov's hypothesis \cite{Sakharov:1967dj}, they can co-generate both ordinary and mirror baryon asymmetries \cite{Bento:2001rc,Bento:2002sj} and thus naturally explain the concordance between dark and ordinary matter fractions in the Universe, $\Omega_{B'}/\Omega_B\simeq 5$ 
\cite{Berezhiani:2008zza,Berezhiani:2018zvs}.

The neutron and mirror-neutron mass mixing $\epsilon_{n n'} +
\mathrm{h.c.}$    
in contrast to the neutron-antineutron mixing \cite{Phillips:2014fgb}, is not severely restricted by 
the experimental limits; they allow an $n-n'$ oscillation time, 
$\tau_{nn'}=\hbar\epsilon^{-1}$, as small as few seconds \cite{Berezhiani:2005hv},  
in agreement also with astrophysical bounds 
from cosmic rays \cite{Berezhiani:2006je,Berezhiani:2011da}
and neutron stars \cite{Berezhiani:2020zck,McKeen:2021jbh,Berezhiani:2021src,Goldman:2022brt}. 
Fast $n-n'$ transitions are not easily detectable as far as they are suppressed by environmental factors  
as the presence of ordinary and/or mirror matter or of ordinary and/or mirror magnetic fields 
\cite{Berezhiani:2005hv,Berezhiani:2009ldq,Berezhiani:2012rq}. 
Various proposals for experimental searches via neutron disappearance $n\to n'$ and regeneration 
$n\to n' \to n/\bar n$, and neutron spin-precession distortion 
can be found in 
\cite{Pokotilovski:2006gq,Berezhiani:2009ldq,Berezhiani:2017azg,Berezhiani:2020vbe,Berezhiani:2018qqw,Kamyshkov:2021kzi}. 
The possible effects for neutron lifetime measurements 
and exotic nuclear transitions were discussed in  \cite{Berezhiani:2018eds,Berezhiani:2018udo,Berezhiani:2015afa}.

The observation that the oscillation time 
$\tau_{nn'}$ may be less than the neutron lifetime~\cite{Berezhiani:2005hv} triggered an intense experimental search for anomalous $n\to n'$ losses
using ultra-cold neutrons (UCNs). 
A series of experiments 
\cite{Ban:2007tp,Serebrov:2007gw,Bodek:2009zz,Serebrov:2008her,Altarev:2009tg,Berezhiani:2017jkn,nEDM:2020ekj}, performed at various magnetic fields testing energy degeneracy, 
established significant upper bounds on $\tau_{nn'}$. 
In a reanalysis of data from ~\cite{Ban:2007tp,Serebrov:2008her} the authors of~\cite{Berezhiani:2012rq,Berezhiani:2017jkn} found significant deviations from the null-hypothesis, namely $3\sigma$ and $5\sigma$ effects. 
The data of 
\cite{Berezhiani:2017jkn} have shown a $2.5\sigma$ deviation.
They could explain the deviations via oscillations in the mass-degenerate $n-n'$ system in the background of a mirror magnetic field, $B'$, from mirror-photons~\cite{Berezhiani:2009ldq,Berezhiani:2012rq}. 
The next storage experiment performed by the nEDM collaboration at PSI~\cite{nEDM:2020ekj,MohanMurthy} has excluded a big part of the parameter space 
for the above deviations but some parameter regions were left. 

In an effort to test the remaining parameter space, a collaboration at PSI assembled a new experiment with a dedicated apparatus described in \cite{Ayres:2021zbh,Ingo}. The measurement series was performed in 2021.
In this Letter we report on the analysis and the obtained exclusion limits. 

In the UCN storage experiments, $n \to  n'$ transitions would manifest as additional neutron losses unaccounted for by other loss channels such as neutron decay or during wall collisions. Since $n'$ is sterile 
against conventional interactions, at each wall collision it would escape the storage chamber and thus elude detection.   
In the mass degenerate neutron $-$ mirror-neutron system, the 
$n \to n'$ transition probability 
as a function of free flight time $t$ between collisions is \cite{Berezhiani:2009ldq,Berezhiani:2012rq}:
\begin{equation}
\begin{split}
P^{nn'}_{BB'}(t) &= \frac{\sin^2[(\omega-\omega')t]}{2\tau^2_{nn'}(\omega-\omega')^2} +  \frac{\sin^2[(\omega+\omega')t]}{2\tau^2_{nn'}(\omega+\omega')^2} \\ &+ \left(\frac{\sin^2[(\omega-\omega')t]}{2\tau^2_{nn'}(\omega-\omega')^2} -  \frac{\sin^2[(\omega+\omega')t]}{2\tau^2_{nn'}(\omega+\omega')^2} \right)\cos \beta .
\end{split}
\label{eq:prob}
\end{equation}
Here $\omega = |\mu_nB|/2\hbar$ and $\omega' = |\mu_nB'|/2\hbar$, and $\beta$ is the angle between the ordinary magnetic field $\textbf{B}$ and the mirror field $\textbf{B}'$ vectors. This equation is valid as long as both fields are homogeneous and $\beta$ stays constant during the time $t$. 

In the limit $B'\to 0$, the last term in Eq.~\eqref{eq:prob} vanishes, 
and the $n-n'$ oscillation probability is suppressed by the ordinary magnetic field, independently of its direction.  
In this case, the most stringent experimental limit  $\tau_{nn'} > \SI{414}{\second}$ (90\% C.L.)  was obtained in \cite{Serebrov:2007gw}.
However, this limit becomes invalid if a large enough mirror field is present on Earth (assumed to be rotating with it~\cite{Berezhiani:2009ldq}),  $B'\gtrsim \SI{0.1}{\micro\tesla}$. 
In this case, the transition probability, Eq.~\eqref{eq:prob}, 
is a function of both magnitude and direction of the applied magnetic field $\textbf{B}$. 
The oscillations will be suppressed at $B=0$ and resonantly amplified for $B \approx B'$. Because the magnitude and direction of the mirror field are unknown, the magnetic field $B$ needs to be scanned. 

Our experiment employs tunable magnetic fields in the range $\SI{5}{\micro\tesla} < B < \SI{109}{\micro\tesla}$.  
We work with the directional asymmetry $A_B$ as the experimental observable \cite{Berezhiani:2009ldq}:
\begin{equation}
    A_B = \frac{n_B^{(t_s)}-n_{-B}^{(t_s)}}{n_B^{(t_s)} + n_{-B}^{(t_s)}} = 
    \frac{\exp(-m_sP^{nn'}_{BB'})- \exp(-m_sP^{nn'}_{-BB'})}
    {\exp(-m_sP^{nn'}_{BB'}) + \exp(-m_sP^{nn'}_{-BB'})}.
\label{eq:asym}
\end{equation}
Here, $n_{\pm B}^{(t_s)}$ represent neutron counts after a storage time $t_s$, when applying magnetic fields of opposite polarities, $\pm B\textbf{e}_z$. With $\textbf{e}_z$ we denote the unit vector along the vertical axis of the storage vessel. $P^{nn'}_{BB'}$ is the mean probability, Eq.~\eqref{eq:prob}, for a time-lapse between wall collisions,  
and $m_s$ is the mean number of collisions of UCNs detected after storage time $t_s$. 
All conventional losses cancel since they are independent of magnetic fields.   

Until now, we described the case where $\textbf{B}$ is homogeneous. However, this strong assumption for $\textbf{B}$ cannot be made in general. Due to the remaining small magnetization of the storage vessel and external field disturbances, there can be an offset in magnitude 
between the fields set with opposite polarities
under the conditions of the experiment described here. One can also have non-negligible transversal field components that do not exhibit rotational symmetry. Due to these components, the projection $\cos{\beta}$ between the local $\textbf{B}$ and $\textbf{B}'$ changes along the UCN trajectory. Assuming $\textbf{B}'$ from Earth is locally homogeneous and static, we parametrize $\textbf{B}'$ by its magnitude $B'$, the polar angle $b$, and the azimuth angle $a$. 
Similarly, we take the ordinary field vector $\textbf{B}$ as a function of a position vector $\textbf{x}$. Then
\begin{equation}
    \textbf{B}' = B'
    \begin{pmatrix}
        \sin{b}\cos{a}\\
        \sin{b}\sin{a}\\
        \cos{b}
    \end{pmatrix}, 
\textbf{B} = B(\textbf{x})
    \begin{pmatrix}
        \sin{\theta(\textbf{x})}\cos{\phi(\textbf{x})}\\
        \sin{\theta(\textbf{x})}\sin{\phi(\textbf{x})}\\
        \cos{\theta(\textbf{x})}
    \end{pmatrix}
\end{equation} 
The local $\cos{\beta}$ projection in Eq.\eqref{eq:prob} thus depends on  $\textbf{x}$: 
\begin{equation}
\begin{aligned}
    \cos{(\beta\textbf{(x}))} & = \sin{b}\cos{a}\sin{\theta}\cos{\phi} \\
    & + \sin{b}\sin{a}\sin{\theta}\sin{\phi} + \cos{b}\cos{\theta}.
\label{eq:angles}
\end{aligned}
\end{equation}

In the case of an inhomogeneous magnetic field, we consider the local increment in the oscillation probability, that is  the derivative of the probability function  from  Eq.~\eqref{eq:prob} at a time $t_i$, and an infinitesimally short duration. The oscillation probability for a single neutron during a single free flight path, $P_f$, after neglecting higher order terms, becomes the sum of all infinitesimal probabilities along the path. 
\begin{comment}\begin{equation}
    P_f = \sum_i \text{d}P_i
\end{equation}
\end{comment}
The oscillation probability of a single neutron for the entire duration of a storage measurement is obtained by summing the probabilities, $P_f$, for all free flight segments. This is equivalent to $m_sP^{nn'}_{BB'}$ in Eq.~\eqref{eq:asym}, or, in Monte Carlo (MC) simulations, one can directly obtain the cumulated probability for the entire time spent by the UCNs in the magnetic field.
In the general case of a rotationally asymmetric inhomogeneous field and an offset when changing the field polarity, Eq.~\eqref{eq:asym} can be expressed as:  
\begin{equation}
\begin{aligned}[b]
\tau_{nn'} &= \frac{1}{\sqrt{2|\langle A_B \rangle|}} |F^0_A(\textbf{B},B') + F^1_A(\textbf{B},B')\cos{b}\\ & + F_A^2(\textbf{B},B')\sin{b}\cos{a} + F_A^3(\textbf{B},B')\sin{b}\sin{a})|^{\frac{1}{2}}.
\label{eq:tau}
\end{aligned}
\end{equation}
The terms $F^i_A(\textbf{B},B')$, referred to later as  "resonance functions", 
represent asymmetries from $\textbf{B}$ and $-\textbf{B}$ contributions to the cumulated oscillation probabilities (considered as $\ll 1$)
and are calculated in MCUCN~\cite{Zsigmond:2017gmi,Bison:2019hot,
Bison:2021rdj} simulations. 
The term $F_A^0$ arises due to the vertical field gradient and offset when changing field polarity. The terms $F_A^1, F_A^2, F_A^3$ also include the rotational dependency of the $\textbf{B}$ vector. The direction of the mirror field $\textbf{B}'$ is unknown, therefore $\tau_{nn'}$ is given as an explicit function of the angles $a$ and $b$.

%% file: experiment.tex
\begin{figure}
    \includegraphics[width=\linewidth]{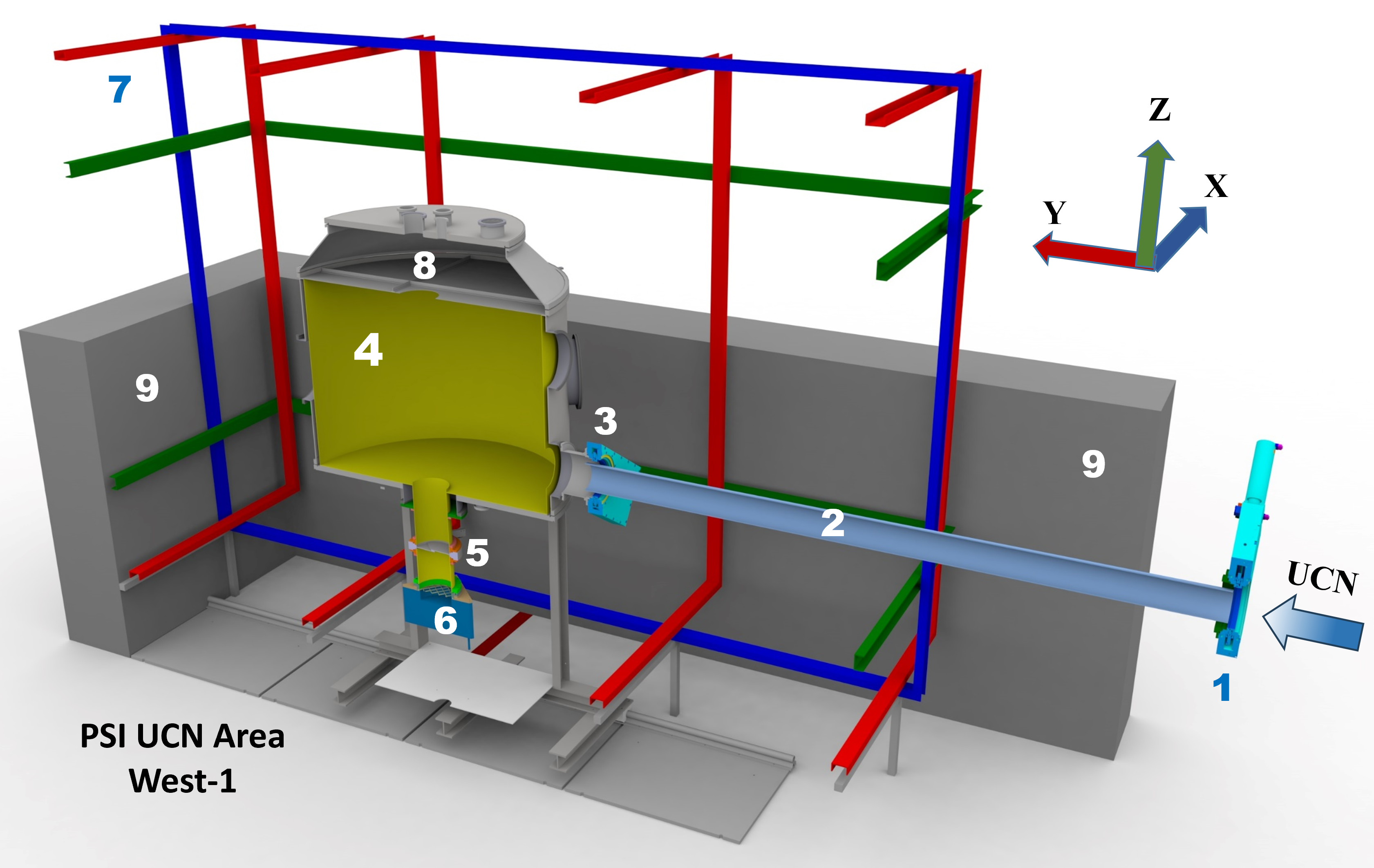}
    \captionsetup{justification=Justified}

    \caption{Cut through the schematic (CAD)  experimental setup: (1) West-1 beamport shutter, (2) UCN guides, (3) horizontal guide shutter, (4) storage vessel, (5) butterfly shutter, (6) UCN detector, (7) magnetic field coils, (8) top lid of the storage vessel, (9) radiation shielding. Figure adapted from \cite{Ingo}.}
    \label{fig:setup}
\end{figure}

%\subsection{Setup}
A schematic view of the apparatus is shown in Fig.~\ref{fig:setup}. It was installed in Area West of the PSI UCN source~\cite{Anghel:2009272,Bison:2019hot,
Bison:2021rdj,Lauss:2021rac} and upgraded in 2021~\cite{Ayres:2021zbh}. UCNs entered through the West-1 beamport shutter  
    (1) and traveled down the Ni/Mo-coated glass guides (2). By passing through the vacuum-tight shutter at the end of the horizontal guide (3) they entered a 1~m high, 1.5 m$^3$ large evacuated storage vessel (4) made of stainless steel. During monitoring and after storage, the neutrons were released through a fast butterfly shutter~\cite{Bison2016449}
    (5) and counted by a 20~cm $\times$ 20~cm CASCADE~\cite{CASCADE2025} UCN detector (6). A set of eight rectangular coils (7), spanning across the entire vacuum tank of the storage vessel was used to generate the target magnetic field. 
    The top lid of the storage vessel (8) could be opened to insert a device to perform spatially resolved measurements of the magnetic field in the storage volume. 
    The setup was surrounded by a wall of concrete blocks for radiation protection (9).

%\subsection{Magnetic field}
The magnetic field was generated by the eight coils wound on aluminium rails arranged 
around the storage vessel (compare also to~\cite{Brys:2005}). Two independent, rectangular coils measuring each $\SI{3}{\meter}\times\SI{4.6}{\meter}$ with distances of $\SI{1.8}{\meter}$ between them produced the fields in $X$- and $Z$-direction. The field along the beamline axis (Y) was generated by two sets of two $\SI{3}{\meter}\times\SI{3}{\meter}$ coils located at distances of $\pm \SI{0.9}{\meter}$ and $\pm \SI{2.2}{\meter}$ from the center of the storage vessel. The two neighbouring Y-coils were driven in series, resulting in a total of six independent currents $\textbf{I} = (I_X^+,I_X^-,I_Y^+,I_Y^-,I_Z^+,I_Z^-)$.
The surrounding field was constantly monitored by 16 three-axis fluxgates mounted on a wooden support structure around the storage vessel (see
details on their sensitivities in~\cite{Ingo}).

To characterize the magnetic field generated by the coils we performed spatially resolved measurements of the magnetic field response to the currents both inside the storage vessel, using a dedicated device, called a mapper, and the aforementioned monitoring fluxgates. 
The mapper (Fig.~\ref{fig:mapper} in Supplemental Material) held five fluxgates, which were mounted on a wood-polypropylene-profile spanning the diameter of the vessel. The profile was attached to a long, hollow aluminum rod that had to be inserted into the vessel from the top. 
The height and rotational angle of the mapper were tracked using potentiometer-based position sensors. 
 
A single map consisted of rotational scans at several heights. To characterize the magnetic field inside the storage vessel as a function of the applied coil currents, at least seven maps at linearly independent current combinations had to be taken. 

Using the method proposed in~\cite{Abel:2018arc} one can describe an arbitrary magnetic field as an infinite sum of harmonic polynomials as
\begin{equation}
    \textbf{B}(\textbf{r}) = \sum_{l, m} G_{l,m} \Pi_{l,m}(\textbf{r}).
    \label{eq:polynomials}
\end{equation}
The expansion coefficients $G_{l,m}$ are the generalized gradients for modes of degree $l$. The harmonic polynomials $\Pi_{l,m}(\bf{r})$ are orthogonal and each basis state satisfies Maxwell's equations. 

By fitting the above field model to the data extracted from the magnetic field maps, we obtained an expression for the magnetic field inside the UCN storage vessel as a function of coil currents and position. Due to the design of the coils, it was sufficient to take into account gradients up to the 2nd order. 
These expansion coefficients served as magnetic field input for simulations with the MCUCN code, which were calibrated with measurements of the UCN counts after different storage periods and corresponding emptying time distributions~\cite{Ingo}.  

\begin{figure}
    \includegraphics[width=\linewidth]{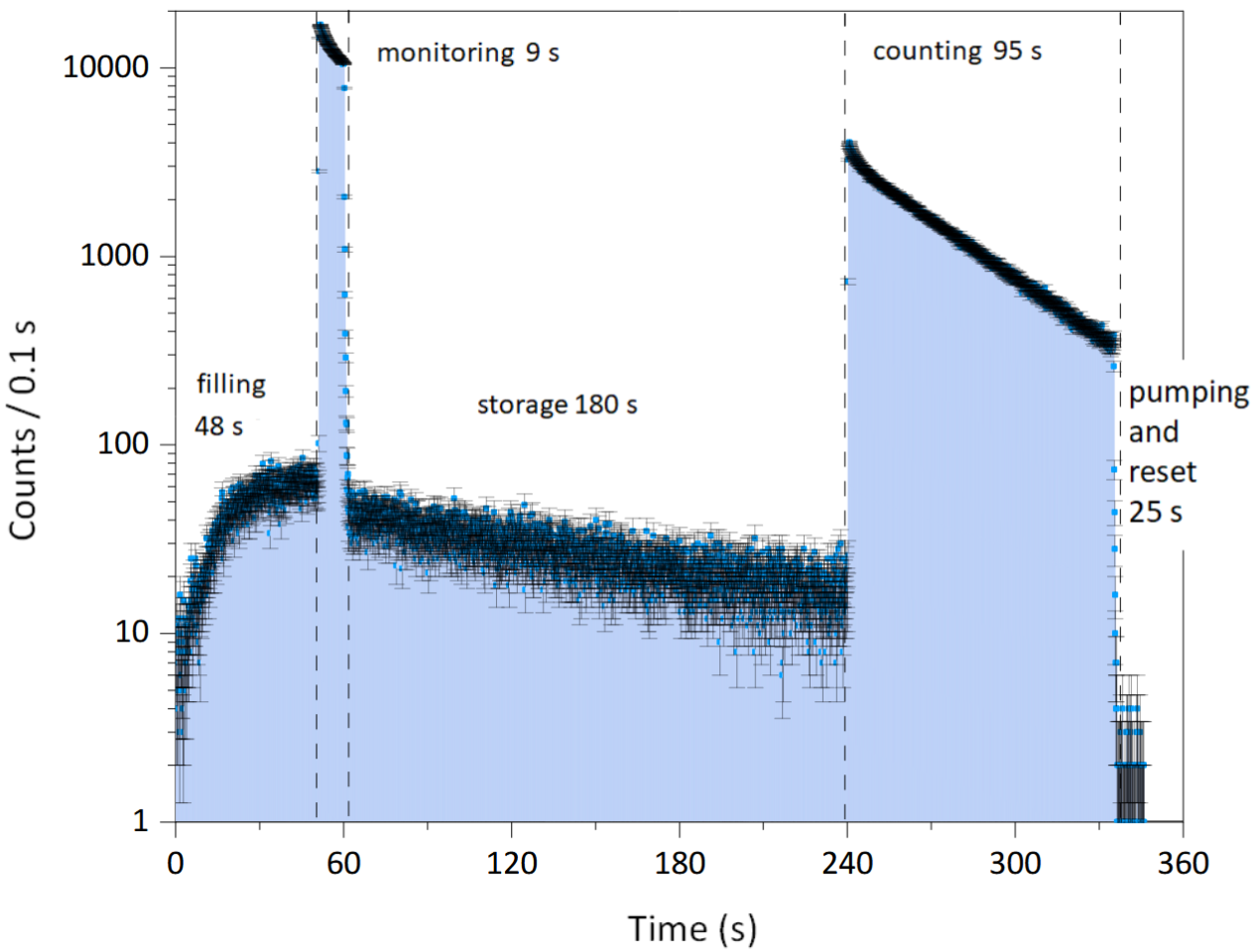}
    \captionsetup{justification=Justified}

    \caption{UCN counts during one storage measurement cycle (adapted from \cite{Ingo}).}
    \label{fig:spectrum}
\end{figure}

A single UCN storage measurement was called a cycle. A detailed description of the order of events happening during one measurement cycle can be found in \cite{Ingo}. The detection rate of the UCN counts as a function of time during a cycle is shown in Fig.~\ref{fig:spectrum}. At $t=0$ a proton beam pulse hit the lead spallation target of the UCN source. The beamport and horizontal guide shutters were open and neutrons entered the storage vessel. During the monitoring period, the butterfly shutter to the detector was briefly opened. After a storage time of $\SI{180}{\second}$, the butterfly shutter was opened again. The UCNs were released and counted for $\SI{95}{\second}$. 

To compensate for long-term drifts in UCN counts, the direction of the applied magnetic field was altered from one cycle to the next. This was done according to the pattern ABBA BAAB, where A ideally corresponds to a target field $\textbf{B} = +|B|\textbf{e}_z$ and B to $\textbf{B} = -|B|\textbf{e}_z$. A complete set of eight cycles was grouped into a sequence. A sequence was the smallest unit of measurements considered in this analysis. 

The UCN DAQ performed 5$\sigma$-clipping of the normalized counts to reject sequences with outliers. These were caused by sequencing errors, proton beam interlocks, UCN source control problems, or synchronization problems with the optical link for the detector readout, leading to clear outliers in the monitoring or storage counts. Data were also rejected in case of problems with the magnetic field control system. 

Strong superconducting magnets from neighbor experiments sometimes caused local field changes of up to several tens of 
microtesla. While the control system of the magnetic field was able to keep the field stable against small outside disturbances, the coil current supplies did not deliver enough current to fully compensate zero-order changes that were caused by these superconducting magnets. The coils were also not designed to counteract changes in gradients higher than second order. The influence of our coils on the field inside the vacuum storage tank was known; however, the effect of external magnetic field changes on the fields inside the vacuum vessel could not be fully characterized. Measurement cycles taken under these conditions were rejected.

%% file: analysis.tex
%\subsection{UCN data analysis}
We calculate the asymmetry between A- and B-cycles for an individual sequence $j$, $A_j$ from
\begin{equation}
    A_j = \frac{\sum_i n_{A,i} - \sum_i n_{B,i}}{\sum_i n_{A,i} + \sum_i n_{B,i}}.
\label{eq:asymmetry}
\end{equation}

\begin{figure}
    \captionsetup{justification=Justified}

    \includegraphics[width=\linewidth]{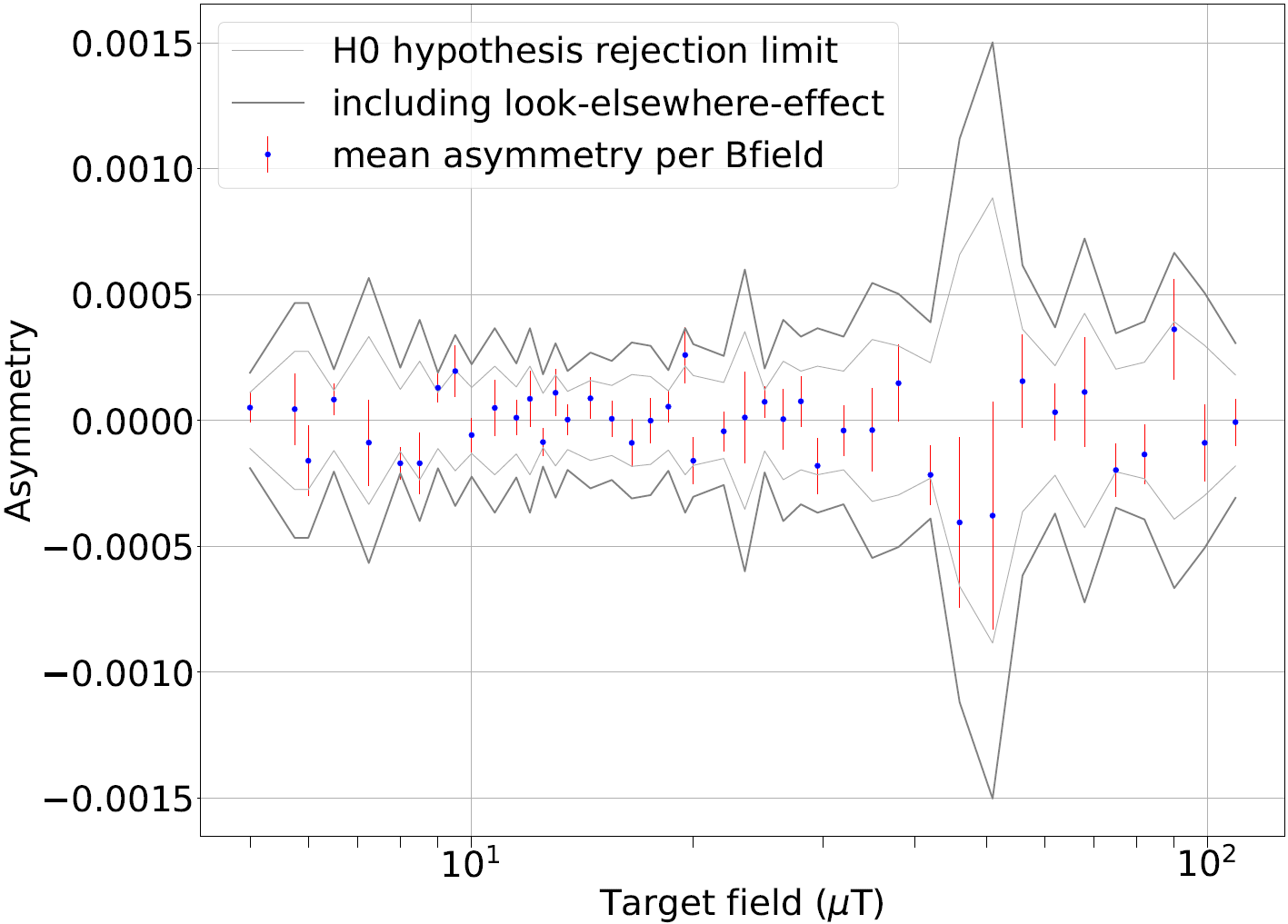}
    \caption{Mean asymmetry $\langle A_B \rangle$ as a function of target field $\textbf{B}$. The light grey line shows the limits of the null-hypothesis $H_0$ at the 95\% C.L., the dark grey outline shows the same limit after correcting for the look-elsewhere-effect. $\langle A_B \rangle$ is consistent with $H_0$ within a standard error $\delta \langle A_B \rangle$.}
    \label{fig:asymmetry}
\end{figure}

The normalized counts $n_\text{\{A,B\},i}$ per cycle ($i=1,2,3,4$) are the ratio of counts after storage and the monitoring counts. The mean asymmetry $\langle A_B \rangle$ per target field  and their standard errors $\delta \langle A_b \rangle$ are plotted in 
Fig.~\ref{fig:asymmetry}, showing that the values of $\langle A_B \rangle$ are consistent with the null-hypothesis in which the measured values of $\langle A_B \rangle$ are compatible with a random sampling of a Gaussian distribution centered at zero and with a standard deviation of the measured uncertainty. Consequently,  no $n-n'$ oscillations were observed, within 95\% C.L. for almost all target fields with or without correcting for the look-elsewhere-effect. After correcting for the look-elsewhere-effect with the Bonferroni procedure~\cite{Dunn:1961}, $n-n'$ oscillations were excluded within 95\% C.L. also at the remaining fields $B = \SI{9}{\micro\tesla}$ and $B = \SI{19.5}{\micro\tesla}$.

%\subsection{Obtaining the limits}

\begin{figure*}

    \includegraphics[width=0.7\linewidth]{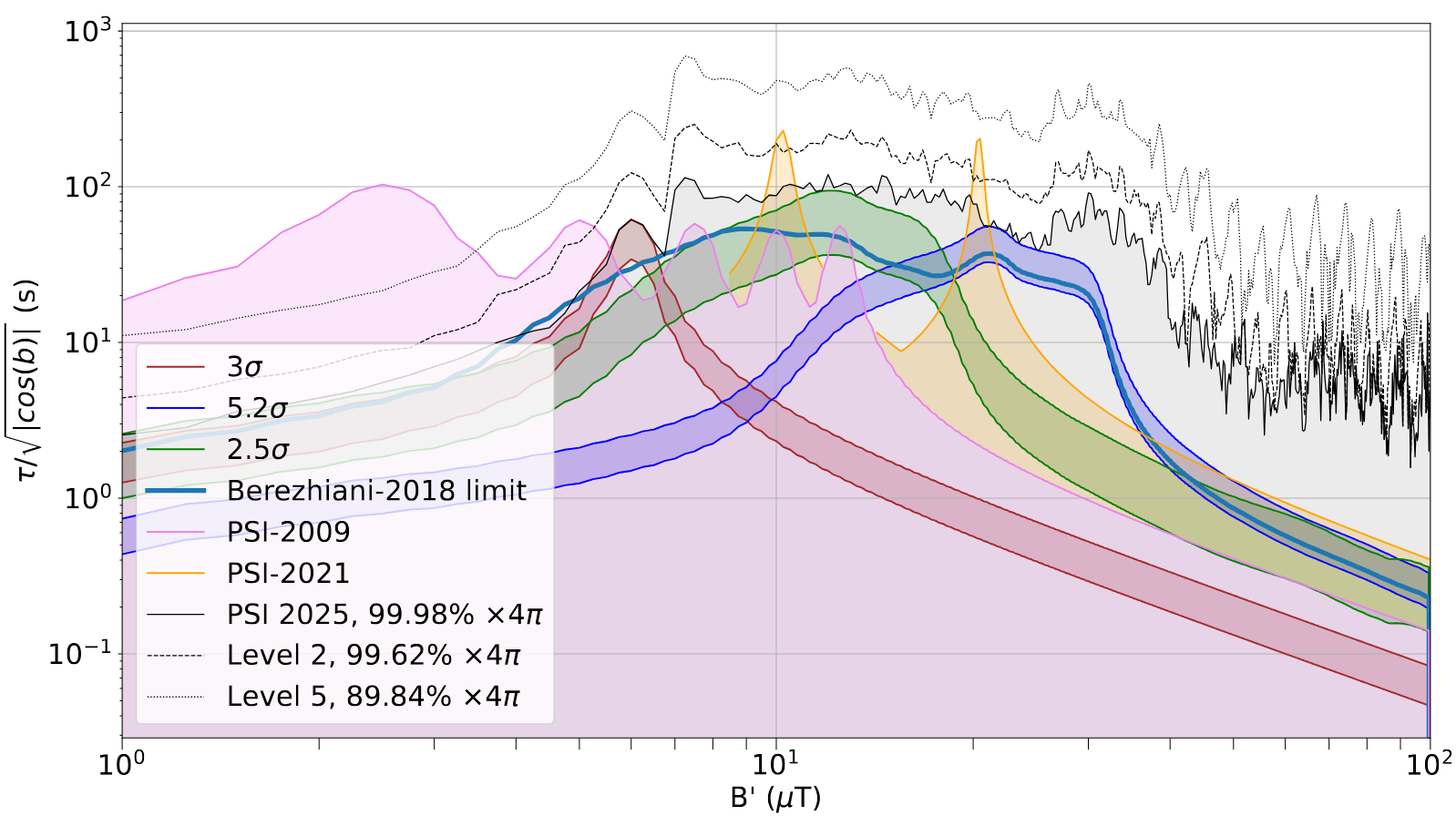}
    \captionsetup{justification=Justified}

    \caption{The anomalous signal regions  of Berezhiani-2018~\cite{Berezhiani:2017jkn} from data in \cite{Ban:2007tp} ($3\sigma$, red),
\cite{Serebrov:2008her} ($5.2\sigma$, blue) and 
\cite{Berezhiani:2017jkn} ($2.5\sigma$, green)
    together with the 95\% C.L. limits on $\tau_{nn'}/\sqrt{|\cos(b)|}$ from PSI-2009~\cite{Altarev:2009tg}, PSI-2021~\cite{nEDM:2020ekj} and this work as a function of the magnitude of the mirror-magnetic-field, $B'$. Along with the most conservative PSI-2025 limit (black line) excluding 99.98\% of the full solid angle (4$\pi$), two higher limit curves but with less solid angle exclusion (99.62\% and 89.84\%) are plotted, corresponding to levels of factor 2 (dashed line) and 5 (dotted line) above the anomalous signals.
    }
    \label{fig:limit_range}
\end{figure*}

We can construct limits for $\tau_{nn'}$ according to Eq.~\eqref{eq:tau} by combining the resonance functions $F_A^0$, $F_A^1$, $F_A^2$, $F_A^3$ we obtained from Monte Carlo simulations of the $n-n'$ oscillation probabilities with the distributions for the asymmetry obtained from the UCN counts. 
From sampling the asymmetry $A_B$ in Eq.~\eqref{eq:tau} as a normal distribution around the measured  mean $\langle A_B \rangle$ with  standard error $\delta \langle A_B \rangle$ from Fig.~\ref{fig:asymmetry}, we obtained the 95\% C.L. limits for $\tau_{nn'}$ as a function of the mirror magnetic field $\textbf{B}'$.
For every target field we obtain one function for $\tau_{nn'}/\sqrt{|\cos(b)|}$, the ratio utilized for the anomalies in~\cite{Berezhiani:2017jkn}. 
Taking the upper envelope of all resonance functions (thus exploiting the best sensitivity from each target field) yields the limit for a given combination of $a$ and $b$. 
When sampling these unknown angular coordinates of the mirror-magnetic-field vector, the majority of the $(a,b)$ tuples provides limits that completely exclude all three anomaly signals, where they were not yet excluded by PSI-2009~\cite{Altarev:2009tg} ($B'>\SI{5.4}{\micro\tesla}$). The lower envelope of these limits
is plotted as "PSI-2025" in Fig.~\ref{fig:limit_range}. Within 99.98\% of the full $4\pi$ solid angle, all previously claimed anomalies are excluded at 95\% C.L. There remains a small region $|\cos(b)|-1< 10^{-4}$ in which these anomalies are not excluded. In this direction of the mirror magnetic field vector, the apparatus had the lowest sensitivity for the oscillation time $\tau_{nn'}/\sqrt{|\cos(b)|}$. The parameter space is three-dimensional. In order to quantify the correlation of the limits $\tau_{nn'}/\sqrt{|\cos(b)|}$ as a function of $B'$ and the angular coordinates, $a$ and $b$, we calculated the limit curves
able to exclude the anomaly signal if it were increased by a factor of 2 and 5 (Fig.~\ref{fig:limit_range}). The exclusion solid angles obtained were 99.62\% and 89.84\% of $4\pi$, respectively. For a detailed mapping of these solid angle regions we refer to the Supplemental Material (Fig.~\ref{fig:heatmap}).

Additional limits obtained by studying 
the frequency variations in the neutron spin-precession 
\cite{Mohanmurthy:2022dbt,MohanMurthy:2024skb} 
regard small values of $B'$ ($< \SI{1}{\micro\tesla}$), outside the range of 
Fig.~\ref{fig:limit_range}.  
Other complementary efforts include the search of $n-n'$ mixing 
 in beam experiments \cite{Ban:2023cja,Broussard:2021eyr,Gonzalez:2024dba,Almazan:2021fvo} 
 but their results are applicable when $n$ and $n'$ states have substantial mass splitting and do not interfere with the parameter space analyzed in our paper.
 
 In conclusion, in a dedicated experiment we have demonstrated that previously claimed anomalies in the $B'$ range of 5 $-$ \SI{109}{\micro\tesla} could be excluded at 95~\% C.L. in 99.98~\% of the 4$\pi$ solid angle of the directional coordinates.

%% file: appendix.tex
\clearpage
\onecolumngrid    
\section{\begin{Large}Supplemental Material\end{Large}}

\hspace{5cm}

\subsection{Magnetic field mapper for the $n - n'$ experiment}

\hspace{5cm}

\begin{figure}[h]
    \includegraphics[width=0.53\linewidth]{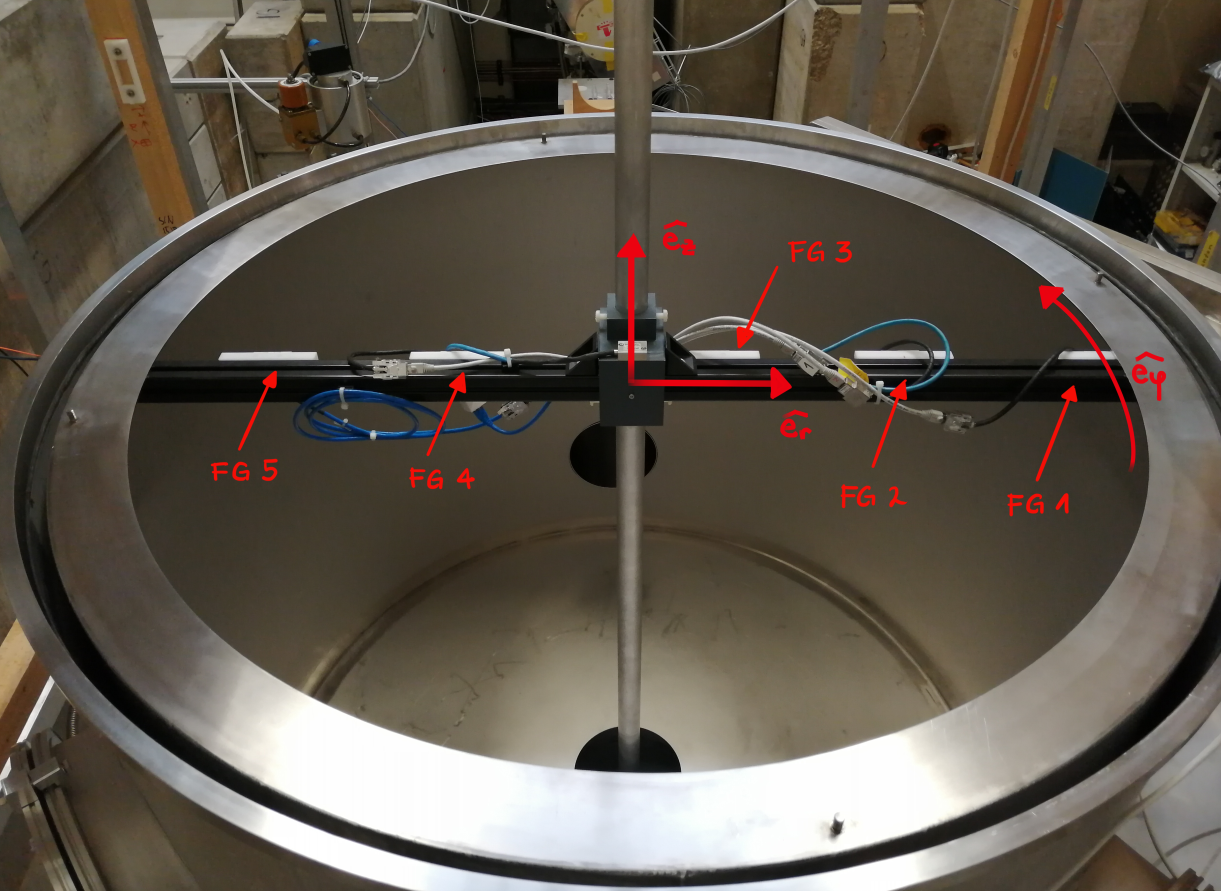}
    \captionsetup{justification=Justified}

    \caption{Picture of the mapper (see text) inside the electro-polished stainless steel UCN storage vessel. }
    \label{fig:mapper}
\end{figure}

A 3-D magnetic mapper, as shown in Fig.~\ref{fig:mapper}, was built in order to characterize in detail the inhomogeneous field configuration. The five three-axis fluxgates of the type FLC3-70 from Stefan Mayer Instruments (see red marks), FG$_i$ are mounted at various positions along the black PEEK-profile. The central axis of the mapper can be lifted and rotated to ensure good coverage of the vessel's inner volume. For further details, see the main text.

\hspace{10cm}

\subsection{Calculations of the excluded solid angle ranges}

\begin{figure}[h]
    \includegraphics[width=0.5\linewidth]{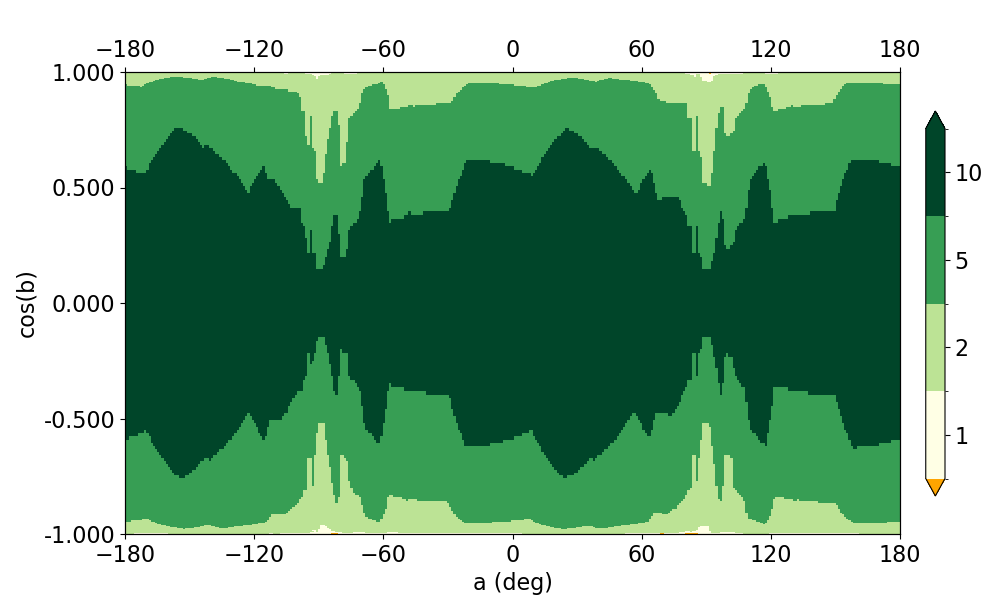}\\
    \includegraphics[width=0.5\linewidth]{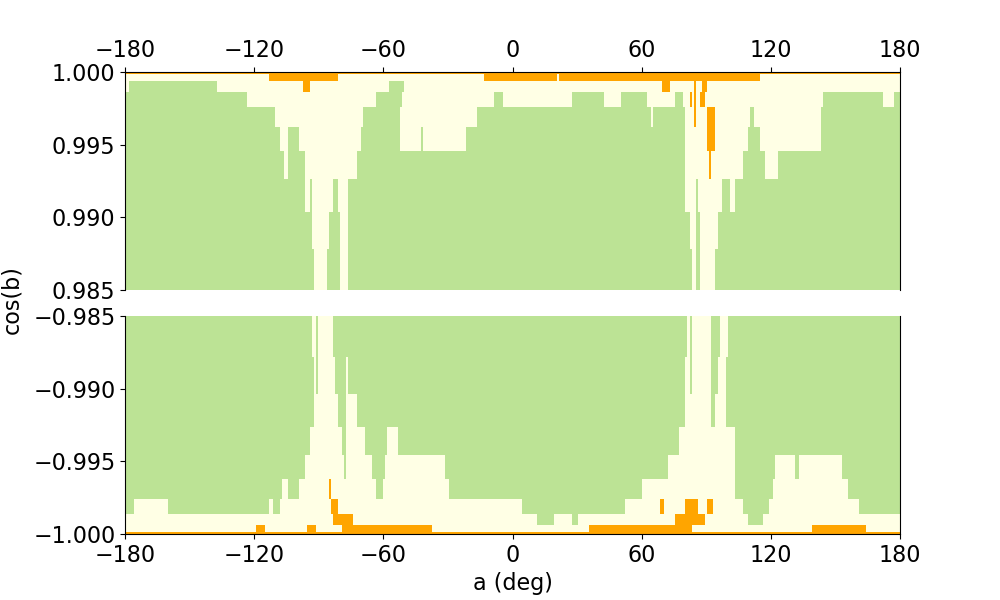}\\
    \captionsetup{justification=Justified}

    \caption{Top $-$ the exclusion regions of the directional parameters, $a$ and $\cos(b)$, of the mirror-magnetic-field vector corresponding to the limit curves from this work shown in Fig.~\ref{fig:limit_range}, see text. 
    Bottom $-$ zoom-in to the solid angle regions at values $|\cos(b)|-1< 10^{-4}$, where the anomalous signals were not excluded (orange on the scale). }
    \label{fig:heatmap}
\end{figure}

To obtain a more general quantitative relation between the height of the exclusion limit in $\tau_{nn'}/\sqrt{|\cos(b)|}$ and the excluded solid angle associated to it, we performed additional calculations by repeating the same analysis on the signal exclusion but setting the signals 'artificially' higher. 
    The anomaly signals were all upscaled by a factor of 2, 5 and 10. 
    The resulting solid angle ranges, in which the signal could still be excluded at 95\% CL by our results, are shown in Figs.~\ref{fig:heatmap} top (full-scale) and bottom (zoom-in).
    
    In Fig.~\ref{fig:heatmap}-Top, the color of each $(a, \cos(b))$ tile indicates the highest factor that the corresponding limit $\tau_{nn'}/\sqrt{|\cos(b)|}$ was able to fully exclude. The  solid angle areas for factors 1, 2, 5 and 10 decrease accordingly as 99.98\%, 99.62\%, 89.84\% and 50.85\% of $4\pi$, respectively. 
    The lack of horizontal symmetry in the patterns is caused by the rotationally non-symmetric field maps that characterize the coil system.
    
    In Fig.~\ref{fig:heatmap}-Bottom, we zoomed-in to the vertical axis values close to $|\cos(b)|-1< 10^{-4}$, to magnify the orange zones, where the anomalous signals were not excluded (compare color scale in Top). The uneven distribution of the orange tiles is caused by the statistical fluctuations in the simulated resonance functions and the discrete sampling of the $(a, \cos(b))$ plane.

